# Equation of Motion Solutions to Hubbard Model retaining Kondo Effect

by

Grzegorz Górski and Jerzy Mizia

Institute of Physics, University of Rzeszów,
ul. Rejtana 16A, 35-958 Rzeszów, Poland
E-mail: ggorski@univ.rzeszow.pl

**Abstract**

We propose a new way of analyzing the Hubbard model using equations of motion (EOM) for the higher-order Green's functions approach within the DMFT scheme. In calculating the higher order Green function we will differentiate over both times ($t$) and ($t'$). This allows us to obtain the metallic Fermi liquid at nonzero Coulomb interaction, where the three center DOS structure with two Hubbard bands and the quasiparticle resonance peak is obtained. At small Coulomb interactions and zero temperature the height of the quasiparticle resonance peak on the Fermi energy is constant similarly as in the full DMFT method with numerical (Quantum Monte Carlo) or with analytical (e.g. iterative perturbation theory) calculations.

PACS codes: 71.10.Fd, 71.27.+a, 71.10.Ay

Keywords: Fermi- liquid, equation of motion approach, Hubbard model

## 1. Introduction

Strongly correlated electron systems are essential for many properties of solids. The basic model describing those systems is the Hubbard model [1, 2], which quantize the relation between the kinetic-energy (itinerant) and potential-energy (localized) effects. Despite its apparent simplicity exact solution of the Hubbard model can be obtained only in some limited cases, e.g. one-dimensional ($d=1$) model [3] and infinite-dimensional ($d=\infty$) model [4]. In the intermediate dimensional systems ($d=2,3,...$) authors use approximate methods or computer simulations. Widely used and accepted is the Hubbard III approximation [2] and corresponding to it the coherent potential approximation (CPA) [5]. This approximations at high Coulomb repulsion $U$, split the spin band into two Hubbard sub-bands: the lower band centered around the atomic level $T_0$ and the upper band centered around the level $T_0 + U$. The defect of classic Hubbard III approximation is that it does not predict any long range ordering, e.g. the ferromagnetic ground state. The magnetism can be brought back by taking into account higher-order correlation function, which generates the spin-dependent shift of the gravity centers of Hubbard sub-bands [6-10]. Hubbard III - like approaches describe well the strong-coupling limits but fail in the weak-coupling limits, in particular they do not reproduce the Fermi-liquid properties for small $U$.

On the other side we have second-order perturbation theory (SOPT) [11, 12], which reproduces well the weak-coupling limits for the case of half-filling. Unfortunately at concentrations away from half-filling the results of this approximation are nonphysical.

Model describing electron correlations in infinite-dimensional systems ($d = \infty$) is the dynamical mean-field theory (DMFT) [13]. It is based on a fact that at $d = \infty$ Hubbard model can be reduced to the single-impurity Anderson model (SIAM) with added self-consistency conditions [14]. Scheme of the DMFT method requires at first step solving the SIAM problem. This can be solved by numerical methods, e.g. quantum Monte Carlo (QMC), exact diagonalization (ED), or numerical renormalization group (NRG), or by approximate analytical methods, e.g. iterative perturbation theory (IPT) [14-18]. Each method has its limitations. To solve the SIAM problem also the equation-of-motion (EOM) approach [19-28] was widely used. The deficiency of the EOM approach until now was not having the Fermi-liquid state at half-filling for $U > 0$ [29]. The reason of this defect was that the EOM schema used strong coupling Green function expansion, which was losing metallic effects, particularly in the particle-hole symmetric case. As was shown by Appelbaum and Penn [21] and Lacroix [19, 20] obtaining the resonance-peak structure in the EOM method is possible only away from the half-filling and in very low temperatures. Temperature increase quickly damped the Kondo peak. The width of this peak was too small. On the other side the EOM approach described well the insulating state at large $U$.

Edwards and Hertz [30] proposed unification of SOPT results with the EOM approach. They started from the alloy-analogy approximation and used second order expansion in $U$ for the self-energy. As was shown by Wermbter and Czycholl [31] the results for infinite-dimensional system (with Gaussian unperturbed density of state and $t^* = 1$) calculated by Edwards-Hertz approximation (EHA) reproduces spectral density with a peak on a Fermi level and the self-energy at small Coulomb interaction $U < 2t^*$. For strong interactions ($U > 4t^*$) and half-filling case the EHA predicts a Mott insulator. For intermediate-coupling strength $2t^* < U < 4t^*$ EHA gives the metallic non-Fermi-liquid phase. Essential weakness of EHA approach is the lack of three peaks at intermediate-coupling strength (lower and upper Hubbard band and the resonance band are not visible) which are present in the DMFT scheme analyzed by QMC, IPT or by us in the current paper.

Irkhin and Zarubin [32] attempted to describe the Fermi-liquid state within the Hubbard III approximation with the use of Hubbard $X$-operators. In their approach the transition to metallic non-Fermi-liquid phase takes place before the metal-insulator transition.

In our paper we will use a modified EOM method to solve the SIAM problem. In calculating the higher order Green function we will differentiate over both times ($t$) and ($t'$). In Section 2 we will describe our modification of the EOM with a better approximation for higher order function $\langle\langle \hat{n}_{d-\sigma} d_\sigma ; d_\sigma^+ \rangle\rangle_\varepsilon$. In Section 3 we will show numerical results of our approach at weak and intermediate Coulomb correlations. Results will be compared with DMFT-IPT and DMFT-VLMA (variational local moment approach) [33] numerical calculations. In Section 4 we present a discussion of the results and our conclusions.

## 2. The model

The Hamiltonian for the single band Hubbard model in a pure itinerant system can be written in the form

$$H = -t \sum_{<ij>\sigma} \left( c_{i\sigma}^+ c_{j\sigma} + h.c. \right) - \mu \sum_i \hat{n}_i + \frac{U}{2} \sum_{i\sigma} \hat{n}_{i\sigma} \hat{n}_{i-\sigma} , \qquad (1)$$

where $t$ is the nearest-neighbors hopping integral, $\mu$ is the chemical potential, $c_{i\sigma}^+ (c_{i\sigma})$ is the operator creating (destroying) an electron with spin $\sigma$ on the $i$-th lattice site, $\hat{n}_{i\sigma} = c_{i\sigma}^+ c_{i\sigma}$ is the electron number operator for electrons with spin $\sigma$ on the $i$-th lattice site, $\hat{n}_i = \hat{n}_{i\sigma} + \hat{n}_{i-\sigma}$ is the operator of the total number of electrons on the $i$-th lattice site. The potential part of the Hamiltonian (1) contains the on-site Coulomb repulsion $U = (i,i|1/r|i,i)$.

The solution to the Hubbard model is the on-site Green function $G_{ii\sigma}(\varepsilon) = \langle\langle c_{i\sigma}; c_{i\sigma}^+ \rangle\rangle_\varepsilon$ given by

$$G_{ii\sigma}(\varepsilon) = \int_{-\infty}^{\infty} \frac{\rho_0(\varepsilon')d\varepsilon'}{\varepsilon + \mu - \varepsilon' - \Sigma_\sigma(\varepsilon)} , \qquad (2)$$

where $\Sigma_\sigma(\varepsilon)$ is the local self-energy and $\rho_0(\varepsilon')$ denotes the free density of states. Now the problem amounts to calculating the quantity $\Sigma_\sigma(\varepsilon)$.

In the DMFT approach we do not solve directly the Hubbard model (1) searching for $\Sigma_\sigma(\varepsilon)$. Instead we are trying to find the solution of the single-impurity Anderson model (SIAM) for which the Hamiltonian may be written as

$$H = \sum_\sigma (\varepsilon_d - \mu)\hat{n}_{d\sigma} + \frac{U}{2}\sum_\sigma \hat{n}_{d\sigma}\hat{n}_{d-\sigma} + \sum_{k\sigma}(\varepsilon_k - \mu)\hat{n}_{k\sigma} + \sum_{k\sigma}\left(V_{dk}d_\sigma^+ c_{k\sigma} + h.c.\right), \qquad (3)$$

where $d_\sigma^+(d_\sigma)$ are the creation (annihilation) operators for the impurity orbital with energy $\varepsilon_d$, $c_{k\sigma}^+(c_{k\sigma})$ are the creation (annihilation) operators for the conduction electron (bath) with the energy dispersion $\varepsilon_k$, $U$ is the on-site Coulomb interaction between electrons on the impurity, and $V_{dk}$ is the coupling between the bath and impurity orbital. We will analyze the paramagnetic case, therefore the spin indices for energy ($\varepsilon_d$, $\varepsilon_k$, $\mu$) and interaction $V_{dk}$ will be neglected.

After solving this model (in different approximations) we obtain the SIAM Green function

$$\langle\langle d_\sigma; d_\sigma^+ \rangle\rangle_\varepsilon = \frac{1}{\varepsilon - \varepsilon_d + \mu - \Delta(\varepsilon) - \Sigma_{d\sigma}} , \qquad (4)$$

with the local self-energy $\Sigma_{d\sigma}$. The hybridization function $\Delta_\sigma(\varepsilon)$ is defined as

$$\Delta_\sigma(\varepsilon) = \sum_k \frac{|V_{dk}|^2}{\varepsilon + \mu - \varepsilon_k} . \qquad (5)$$

Following the DMFT scheme we impose the self-consistency condition demanding that the two Green functions given by eq. (2) and eq. (4) are identical by iterating self-energy until $\Sigma_\sigma(\varepsilon) = \Sigma_{d\sigma}$.

In our analysis we will use the equation of motion for Green functions method to solve the SIAM model. In general the EOM may be written as

$$\varepsilon\langle\langle A;B \rangle\rangle_\varepsilon = \langle[A,B]_+\rangle + \langle\langle[A,H]_-;B\rangle\rangle_\varepsilon . \qquad (6)$$

Applying it to the function $\langle\langle d_\sigma; d_\sigma^+ \rangle\rangle_\varepsilon$ we obtain:

$$\varepsilon\langle\langle d_\sigma; d_\sigma^+ \rangle\rangle_\varepsilon = 1 + (\varepsilon_d - \mu)\langle\langle d_\sigma; d_\sigma^+ \rangle\rangle_\varepsilon + \sum_k V_{kd}\langle\langle c_{k\sigma}; d_\sigma^+ \rangle\rangle_\varepsilon + U\langle\langle \hat{n}_{d-\sigma}d_\sigma; d_\sigma^+ \rangle\rangle_\varepsilon . \qquad (7)$$

For the function $\langle\langle c_{k\sigma}; d_\sigma^+ \rangle\rangle_\varepsilon$ we have

$$\langle\langle c_{k\sigma}; d_\sigma^+ \rangle\rangle_\varepsilon = \frac{V_{dk}}{\varepsilon + \mu - \varepsilon_k}\langle\langle d_\sigma; d_\sigma^+ \rangle\rangle_\varepsilon . \qquad (8)$$

Inserting eq. (8) to eq. (7) we arrive at

$$[\varepsilon - \varepsilon_d + \mu - \Delta_\sigma(\varepsilon)]\langle\langle d_\sigma; d_\sigma^+ \rangle\rangle_\varepsilon = 1 + U\langle\langle \hat{n}_{d-\sigma}d_\sigma; d_\sigma^+ \rangle\rangle_\varepsilon . \qquad (9)$$

Equation (9) can be expressed by

$$[\varepsilon-\varepsilon_d+\mu-\Delta_\sigma(\varepsilon)-Un_{d-\sigma}]\langle\langle d_\sigma;d_\sigma^+\rangle\rangle_\varepsilon = 1+U\langle\langle(\hat{n}_{d-\sigma}-n_{d-\sigma})d_\sigma;d_\sigma^+\rangle\rangle_\varepsilon,\qquad(10)$$

where $n_{d-\sigma}=\langle\hat{n}_{d-\sigma}\rangle$.

To solve eq. (10) we have to write the EOM for the higher order Green function $\langle\langle(\hat{n}_{d-\sigma}-n_{d-\sigma})d_\sigma;d_\sigma^+\rangle\rangle_\varepsilon$. Instead of eq. (6) we will use the method with differentiating over the second time ($t'$), which gives the EOM in the following form [34]:

$$-\varepsilon\langle\langle A;B\rangle\rangle_\varepsilon = -\langle[A,B]_+\rangle+\langle\langle A;[B,H]_-\rangle\rangle_\varepsilon,\qquad(11)$$

from which we obtain

$$[\varepsilon-\varepsilon_d+\mu-\Delta_\sigma(\varepsilon)]\langle\langle(\hat{n}_{d-\sigma}-n_{d-\sigma})d_\sigma;d_\sigma^+\rangle\rangle_\varepsilon = U\langle\langle(\hat{n}_{d-\sigma}-n_{d-\sigma})d_\sigma;(\hat{n}_{d-\sigma}-n_{d-\sigma})d_\sigma^+\rangle\rangle_\varepsilon.\qquad(12)$$

The above equation may be casted into the form:

$$[\varepsilon-\varepsilon_d+\mu-\Delta_\sigma(\varepsilon)-Un_{d-\sigma}]\langle\langle(\hat{n}_{d-\sigma}-n_{d-\sigma})d_\sigma;d_\sigma^+\rangle\rangle_\varepsilon$$
$$= U\langle\langle(\hat{n}_{d-\sigma}-n_{d-\sigma})d_\sigma;(\hat{n}_{d-\sigma}-n_{d-\sigma})d_\sigma^+\rangle\rangle_\varepsilon.\qquad(13)$$

To simplify the notation we introduce the definition

$$\Gamma_{d\sigma}(\varepsilon)=\langle\langle(\hat{n}_{d-\sigma}-n_{d-\sigma})d_\sigma;(\hat{n}_{d-\sigma}-n_{d-\sigma})d_\sigma^+\rangle\rangle_\varepsilon.\qquad(14)$$

Using eqs (10) and (13) and after some algebraic operations we arrive at the expression for the one-particle Green function:

$$\langle\langle d_\sigma;d_\sigma^+\rangle\rangle_\varepsilon = \frac{1}{\varepsilon-\varepsilon_d+\mu-\Delta(\varepsilon)-Un_{d-\sigma}-\dfrac{U^2\Gamma_{d\sigma}(\varepsilon)}{1+U^2\Gamma_{d\sigma}(\varepsilon)G_{d\sigma}^{HF}(\varepsilon)}},\qquad(15)$$

where $G_{d\sigma}^{HF}(\varepsilon)$ is given as

$$G_{d\sigma}^{HF}(\varepsilon) = \frac{1}{\varepsilon-\varepsilon_d+\mu-\Delta_\sigma(\varepsilon)-Un_{d-\sigma}}.\qquad(16)$$

Comparing eqs (4) and (15) we can write for the self-energy:

$$\Sigma_{d\sigma}(\varepsilon)=Un_{d-\sigma}+\frac{U^2\Gamma_{d\sigma}(\varepsilon)}{1+U^2\Gamma_{d\sigma}(\varepsilon)G_{d\sigma}^{HF}(\varepsilon)}.\qquad(17)$$

Expression (17) is the exact formula for the self-energy, but it requires the calculation of the function $\Gamma_{d\sigma}(\varepsilon)$, which leads to higher order equations. We calculate this function in the Appendix using the Kuzemsky Green function decoupling [35] and a well known spectral theorem [36, 37]. Using these approximations we arrive at

$$\Gamma_{d\sigma}(\varepsilon)\approx -\iiint\frac{S_{-\sigma}(x)S_{-\sigma}(y)S_\sigma(z)}{\varepsilon+x-y-z+i0^+}[f(x)f(-y)f(-z)+f(-x)f(y)f(z)]dxdydz,\qquad(18)$$

where the spectral density from eq. (4) can be expressed as

$$S_\sigma(\varepsilon) = -\frac{1}{\pi}\operatorname{Im}G_{d\sigma}(\varepsilon) = -\frac{1}{\pi}\operatorname{Im}\frac{1}{\varepsilon+\mu-\varepsilon_d-\Delta_\sigma(\varepsilon)-\Sigma_{d\sigma}(\varepsilon)}.\qquad(19)$$

## 3. Numerical Results

In this section we will investigate the model numerically. We will use the semi-elliptic DOS corresponding to the Bethe lattice

$$\rho_0(\varepsilon)=\frac{2}{\pi D}\sqrt{1-\left(\frac{\varepsilon}{D}\right)^2},\qquad(20)$$

where $D$ is the unperturbed half bandwidth, which will be assumed as the unit of energy ($D \equiv 1$). Using definition (2) it is easy to show that the unperturbed Green function corresponding to this DOS is given by

$$G_{ii\sigma}^0(\varepsilon) = \frac{2}{D}\left[\frac{\varepsilon}{D} - \sqrt{\left(\frac{\varepsilon}{D}\right)^2 - 1}\right]. \tag{21}$$

This density of states is convenient for the DMFT scheme since it gives a simple relationship between the hybridization function $\Delta_\sigma(\varepsilon)$ and the perturbed Green function $G_{ii\sigma}(\varepsilon)$. To obtain this relationship we will use the expression $G_{ii\sigma}(\varepsilon) = G_{ii\sigma}^0(\varepsilon + \mu - \Sigma_\sigma)$ together with eq. (21), which brings about

$$\varepsilon + \mu - \Sigma_\sigma = \frac{D^2}{4}G_{ii\sigma}(\varepsilon) + G_{ii\sigma}^{-1}(\varepsilon). \tag{22}$$

From relationship (4) we have

$$G_{d\sigma}^{-1}(\varepsilon) + \Delta_\sigma(\varepsilon) = \varepsilon - \varepsilon_d + \mu - \Sigma_{d\sigma}. \tag{23}$$

Assuming $\varepsilon_d = 0$ and the basic DMFT identifications: $G_{ii\sigma}(\varepsilon) = G_{d\sigma}(\varepsilon)$, $\Sigma_{d\sigma} = \Sigma_\sigma$, we obtain that the hybridization function $\Delta_\sigma(\varepsilon)$ is simply related to the Green's function by the relationship

$$\Delta_\sigma(\varepsilon) = \frac{D^2}{4}G_{ii\sigma}(\varepsilon) \equiv \frac{D^2}{4}G_{ii\sigma}^0(\varepsilon + \mu - \Sigma_\sigma). \tag{24}$$

To calculate the spectral density $S_\sigma(\varepsilon)$ we will use a iteration procedure characteristic for the DMFT [13]. We start iteration from any simple $\Sigma_\sigma$, e.g. $\Sigma_\sigma \approx Un_{-\sigma}$. Next we calculate the hybridization function $\Delta_\sigma(\varepsilon)$ from eq. (24) and the spectral density $S_\sigma(\varepsilon)$ from eq. (19). Using this $S_\sigma(\varepsilon)$ we find the function $\Gamma_{d\sigma}(\varepsilon)$ from eq. (18). Having initial function $\Gamma_{d\sigma}(\varepsilon)$ and the hybridization function $\Delta_\sigma(\varepsilon)$ we calculate new self-energy $\Sigma_{d\sigma}$ from eq. (17). Assuming equality between bath and impurity self energy ($\Sigma_\sigma = \Sigma_{d\sigma}$) we come back to eqs (24), (19), (18) and we iterate until the convergence is reached.

In Fig. 1 we present the spectral density of states $S_\sigma(\varepsilon)$ for different $U$ at the half-filling point. Calculations were performed at small value of $U$ characteristic for weak correlation, therefore modification of the noninteracting DOS is only slight. The DOS on a Fermi level is the same as in the unperturbed state $S_\sigma(\mu) = \rho_0(\mu)$ for all $U$. At $U = 1.1$ we see the beginning of the formation of satellite peaks. At stronger correlations ($U > 1.2$) self-energy from eq. (17) with the function $\Gamma_{d\sigma}(\varepsilon)$ given by eq. (18) is not convergent. The reason for this divergence is the denominator $\left[1 + U^2\Gamma_{d\sigma}(\varepsilon)G_\sigma^{HF}(\varepsilon)\right]^{-1}$, which at larger $U$ and $\Gamma_{d\sigma}(\varepsilon)$ calculated from eq. (18) creates singularity.

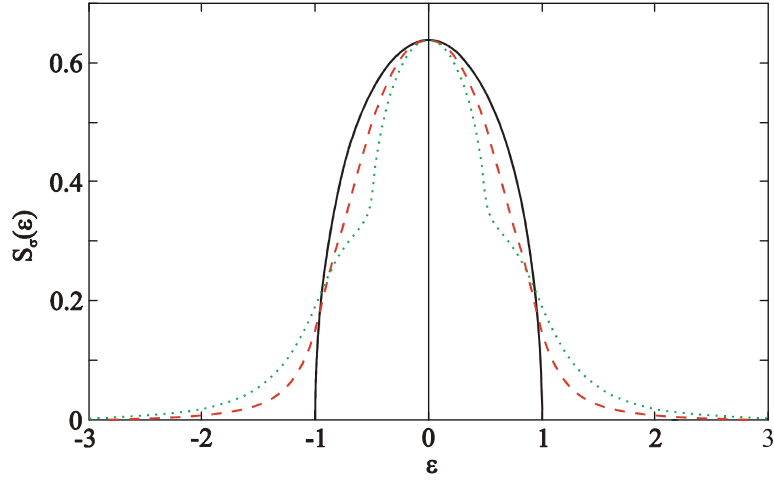

Fig. 1 Spectral density $S_\sigma(\varepsilon)$ as a function of energy $\varepsilon$ for different values of Coulomb interaction at half-filled band and $T=0$. $U=0$ - solid black line, $U=0.7$ - dashed red line, $U=1.1$ - dotted green line. Calculations are based on eq. (17) for the self-energy.

Singularities in the denominator of eq. (17) can be eliminated by calculating $\Gamma_{d\sigma}(\varepsilon)$ in approximations higher than that of eq. (18). Another way of eliminating singularities is to retain $\Gamma_{d\sigma}(\varepsilon)$ of eq. (18) but replace the function $G_\sigma^{HF}(\varepsilon)$ by the Edwards-Hertz [30] propagator $\tilde{G}_{d\sigma}^{HF}(\varepsilon)$ which brings about the expression:

$$\Sigma_{d\sigma}(\varepsilon) = Un_{d-\sigma} + \frac{U^2 \Gamma_{d\sigma}(\varepsilon)}{1+U^2 \Gamma_{d\sigma}(\varepsilon)\tilde{G}_{d\sigma}^{HF}(\varepsilon)} \quad , \tag{25}$$

with the Edwards-Hertz [30] propagator $\tilde{G}_{d\sigma}^{HF}(\varepsilon)$ given as

$$\tilde{G}_{d\sigma}^{HF}(\varepsilon) = \int_{-\infty}^{\infty} \frac{\tilde{\rho}(\varepsilon')d\varepsilon'}{\varepsilon - Un_{d-\sigma} + \mu_{EH} - \varepsilon'} \quad , \tag{26}$$

and

$$\tilde{\rho}(\varepsilon) \approx \frac{1}{n_{-\sigma}(1-n_{-\sigma})} \operatorname{Im} \iiint \frac{S_{-\sigma}^{HF}(x)S_{-\sigma}^{HF}(y)S_\sigma^{HF}(z)}{\varepsilon + x - y - z + i0^+} \times [f(x)f(-y)f(-z) + f(-x)f(y)f(z)]dxdydz \quad , \tag{27}$$

where

$$S_\sigma^{HF}(\varepsilon) = -\frac{1}{\pi}\operatorname{Im}\int_{-\infty}^{\infty} \frac{\rho_0(\varepsilon')d\varepsilon'}{\varepsilon + \mu_{EH} - Un_{d-\sigma} - \varepsilon'} \quad . \tag{28}$$

Parameter $\mu_{EH}$ is fitted to the carrier concentration. As was shown by Wermbter and Czycholl [31] the use of propagator $\tilde{G}_{d\sigma}(\varepsilon)$ in alloy analogy equations allows us to obtain the Fermi liquid behavior and quasiparticle peak at the chemical potential for sufficiently strong Coulomb interactions.

In Fig. 2 we present DOS for several values of the Coulomb correlation calculated with the Edwards-Hertz propagator $\tilde{G}_{d\sigma}^{HF}(\varepsilon)$ in eq. (17). The resultant DOS has a three peaks structure, a resonance peak plus two Hubbard peaks. An increase of $U$ decreases the width of the quasiparticle resonance peak on the Fermi energy and slightly shifts the Hubbard bands to higher energies.

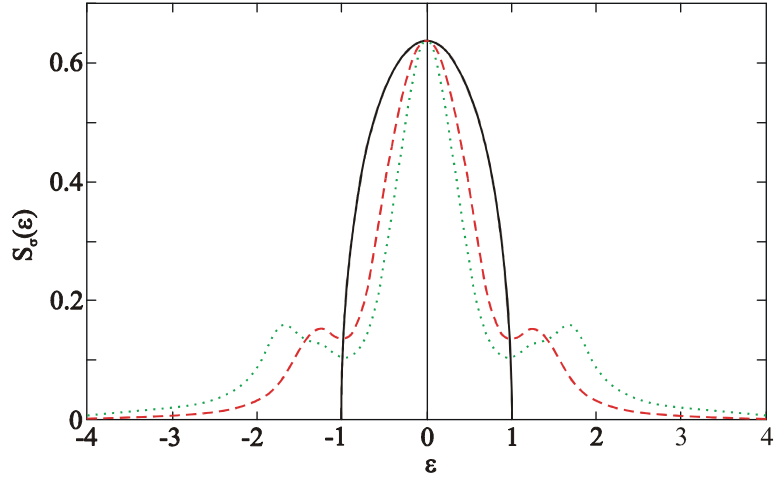

Fig. 2 Spectral density $S_\sigma(\varepsilon)$ as a function of energy $\varepsilon$ for different values of Coulomb interaction at half-filled band and $T=0$. $U=0$ - solid black line, $U=1.5$ - dashed red line, $U=2.5$ - dotted green line. Calculations are based on eq. (25) for the self-energy with Edwards-Hertz propagator $\tilde{G}^{HF}_{d\sigma}(\varepsilon)$.

In Figs 3 and 4 we have compared our results with the IPT approach [16,17] (dashed line) and the Variational Local Moment Approach (VLMA) calculations of Kauch and Byczuk [33] (dotted line). Data in Fig. 3 are calculated using the self energy of eq. (17) and the Coulomb repulsion $U=1$ (in units of half bandwidth). Our results are similar to those obtained by the IPT and VLMA methods. All these approaches have the resonance peak at Fermi energy with the DOS on Fermi energy equal to the DOS of the noninteracting system ($S_\sigma(\mu) = \rho_0(\mu)$) at $T=0$. In Fig. 4 we compare our method with the IPT and VLMA methods for $U=2.4$ using the self energy of eq. (25) with the Edwards-Hertz propagator $\tilde{G}^{HF}_{d\sigma}(\varepsilon)$. In comparison with other methods (IPT and VLMA) the width of the resonance peak in our approach is the largest, and the Hubbard bands are too small. This is the result of the approximating function $\Gamma_{d\sigma}(\varepsilon)$ by the weak coupling expression.

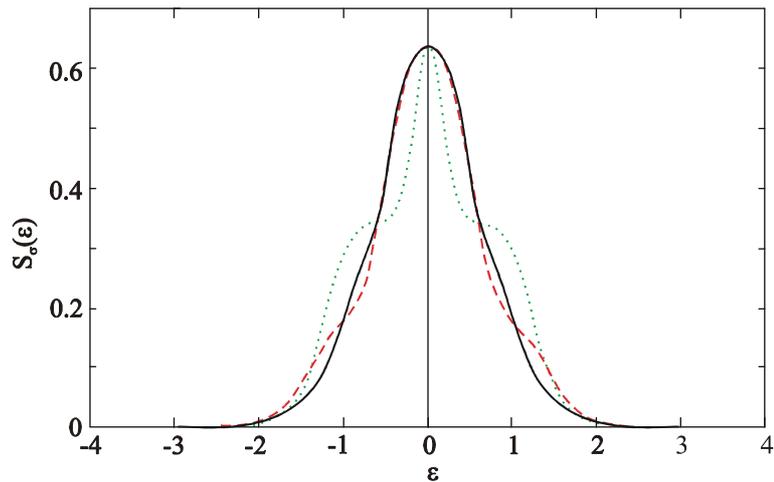

Fig. 3 Spectral density $S_\sigma(\varepsilon)$ as a function of energy $\varepsilon$ calculated by different methods: our self-energy (17) with Green function $G^{HF}_{d\sigma}(\varepsilon)$ – solid black line, IPT approach – dashed red line, VLMA calculations (by Kauch and Byczuk [33]) – dotted green line. The case is that of half-filled band, $U=1$ and $T=0$.

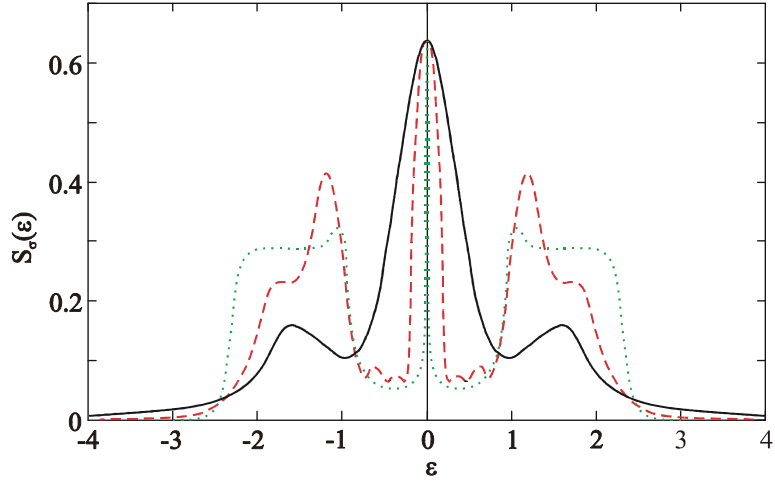

Fig. 4 Spectral density $S_\sigma(\varepsilon)$ as a function of energy $\varepsilon$ calculated by different methods: our self-energy (25) with Edwards-Hertz propagator $\tilde{G}^{HF}_{d\sigma}(\varepsilon)$ – solid black line, IPT approach – dashed red line, VLMA calculations (by Kauch and Byczuk [33]) – dotted green line. The case is that of half-filled band, $U = 2.4$ and $T = 0$.

**4. Conclusions**

In this work we propose a new EOM approach as the solution within the dynamical mean field theory. This method, in a half-filled band at $U \neq 0$, allows us to obtain DOS on the Fermi level to be the same as for the noninteracting system ($S_\sigma(\mu) = \rho_0(\mu)$ for $T = 0$). The density of states has a pronounced three centers structure, which are two Hubbard centers and a quasiparticle resonance peak. This result is an improvement on other EOM methods based on the Appelbaum and Penn [21] and Lacroix [19, 20] approach which does not have the resonance peak on the Fermi energy at half-filled band at finite Coulomb interaction.

Our results are compared with the results of other approaches (IPT and VLMA) at weak ($U = 1$) and intermediate ($U = 2.4$) Coulomb correlations. At weak Coulomb correlation our results are comparable with results of IPT and VLMA. At the intermediate $U$ the width of the quasiparticle resonance peak obtained by us is larger than in the other two methods. The results of the EOM method can be improved at intermediate $U$ by using a higher order approximation for the Green function $\Gamma_{d\sigma}(\varepsilon)$. There is also the question whether coefficient $D^2/4$ in eq. (24) should not be reduced, as at its current value it implies an interaction between the impurity state and all states $\varepsilon_k$ of the bath.

Despite its shortcomings the proposed method has the advantage of being semi analytic and simple for the later applications.

It will be interesting to apply our new method to analyze the ferromagnetic state. Also interesting will be the investigation of the antiferromagnetic state, and subsequently to study its coexistence with the superconducting phase in the high temperature superconductors. This will be the subject of our future work.

**Appendix**

Function $\Gamma_{d\sigma}(\varepsilon)$ can be written as:

$$\Gamma_{d\sigma}(\varepsilon) = \langle\langle \hat{n}_{d-\sigma} d_\sigma ; \hat{n}_{d-\sigma} d_\sigma^+ \rangle\rangle_\varepsilon - n_{d-\sigma} \langle\langle \hat{n}_{d-\sigma} d_\sigma ; d_\sigma^+ \rangle\rangle_\varepsilon \\ - n_{d-\sigma} \langle\langle d_\sigma ; \hat{n}_{d-\sigma} d_\sigma^+ \rangle\rangle_\varepsilon + n_{d-\sigma}^2 \langle\langle d_\sigma ; d_\sigma^+ \rangle\rangle_\varepsilon \quad (A.1)$$

For Green functions $\langle\langle \hat{n}_{d-\sigma}d_\sigma; d_\sigma^+\rangle\rangle_\varepsilon$ and $\langle\langle d_\sigma; \hat{n}_{d-\sigma}d_\sigma^+\rangle\rangle_\varepsilon$ we use the approximation:

$$\langle\langle \hat{n}_{d-\sigma}d_\sigma; d_\sigma^+\rangle\rangle_\varepsilon \approx n_{d-\sigma}\langle\langle d_\sigma; d_\sigma^+\rangle\rangle_\varepsilon \text{ and } \langle\langle d_\sigma; \hat{n}_{d-\sigma}d_\sigma^+\rangle\rangle_\varepsilon \approx n_{d-\sigma}\langle\langle d_\sigma; d_\sigma^+\rangle\rangle_\varepsilon . \quad (A.2)$$

Function $\langle\langle \hat{n}_{d-\sigma}d_\sigma; \hat{n}_{d-\sigma}d_\sigma^+\rangle\rangle_\varepsilon$ can be written as

$$\langle\langle \hat{n}_{d-\sigma}d_\sigma; \hat{n}_{d-\sigma}d_\sigma^+\rangle\rangle_\varepsilon = \int_{-\infty}^{\infty} d(t-t')[-\Theta(t-t')] \\ \times \langle \left[ d_{-\sigma}^+(t)d_{-\sigma}(t)d_\sigma(t), d_{-\sigma}^+(t')d_{-\sigma}(t')d_\sigma^+(t') \right]_+ \rangle \exp\left[ i\frac{\varepsilon}{\hbar}(t-t') \right] \quad (A.3)$$

where $\Theta(t-t')$ is the step function

$$\Theta(t-t') = \frac{i}{2\pi} \int_{-\infty}^{\infty} dx \frac{e^{-ix(t-t')}}{x+i0^+} . \quad (A.4)$$

The mean value in the first term of anticommutator in eq. (A.3) is approximated as

$$\langle d_{-\sigma}^+(t)d_{-\sigma}(t)d_\sigma(t)d_{-\sigma}^+(t')d_{-\sigma}(t')d_\sigma^+(t') \rangle \approx \langle d_{-\sigma}^+(t)d_{-\sigma}(t) \rangle \langle d_{-\sigma}^+(t')d_{-\sigma}(t') \rangle \langle d_\sigma(t)d_\sigma^+(t') \rangle \\ + \langle d_{-\sigma}^+(t)d_{-\sigma}(t') \rangle \langle d_{-\sigma}(t)d_{-\sigma}^+(t') \rangle \langle d_\sigma(t)d_\sigma^+(t') \rangle \quad (A.5)$$

and similarly for the second term.

For the time averages above we use the expressions

$$\langle A^+(t')B(t) \rangle = \frac{1}{\hbar} \int_{-\infty}^{\infty} d\varepsilon' \frac{S_{AB}(\varepsilon')}{e^{\beta\varepsilon'}+1} \exp\left[ -\frac{i}{\hbar}\varepsilon'(t-t') \right], \quad (A.6)$$

and

$$\langle B(t)A^+(t') \rangle = \frac{1}{\hbar} \int_{-\infty}^{\infty} d\varepsilon' \frac{S_{AB}(\varepsilon')e^{\beta\varepsilon'}}{e^{\beta\varepsilon'}+1} \exp\left[ -\frac{i}{\hbar}\varepsilon'(t-t') \right], \quad (A.7)$$

where $S_{AB}(\varepsilon') = -\frac{1}{\pi} Im\langle\langle B, A^+ \rangle\rangle_{\varepsilon'}$.

For averages with $t=t'$ we have:

$$\langle d_{-\sigma}^+(t)d_{-\sigma}(t) \rangle = n_{d-\sigma} . \quad (A.8)$$

Using these approximations in the function $\langle\langle \hat{n}_{d-\sigma}d_\sigma; \hat{n}_{d-\sigma}d_\sigma^+\rangle\rangle_\varepsilon$ we arrive at

$$\langle\langle \hat{n}_{d-\sigma}d_\sigma; \hat{n}_{d-\sigma}d_\sigma^+\rangle\rangle_\varepsilon \approx n_{d-\sigma}^2 G_{d\sigma}(\varepsilon) + \iiint \frac{S_{-\sigma}(x)S_{-\sigma}(y)S_\sigma(z)}{\varepsilon + x - y - z + i0^+} \\ \times [f(x)f(-y)f(-z) + f(-x)f(y)f(z)] dxdydz \quad (A.9)$$

where the spectral density can be expressed from eq. (4) as

$$S_\sigma(\varepsilon) = -\frac{1}{\pi} Im\, G_{d\sigma}(\varepsilon) = -\frac{1}{\pi} Im \frac{1}{\varepsilon + \mu - \varepsilon_d - \Delta_\sigma(\varepsilon) - \Sigma_{d\sigma}(\varepsilon)} . \quad (A.10)$$

By means of eqs (A.1), (A.10) and (A.2) we obtain a final expression for the function $\Gamma_{d\sigma}(\varepsilon)$:

$$\Gamma_{d\sigma}(\varepsilon) \approx \iiint \frac{S_{-\sigma}(x)S_{-\sigma}(y)S_\sigma(z)}{\varepsilon + x - y - z + i0^+} [f(x)f(-y)f(-z) + f(-x)f(y)f(z)] dxdydz , \quad (A.11)$$

which is used in eq. (12) of the paper.